\documentclass[aps,prl,twocolumn,amsmath,amssymb,superscriptaddress]{revtex4}
\usepackage{graphicx}
\usepackage{graphics}
\usepackage{bm}
\usepackage[usenames]{color}
\usepackage{colordvi}
\usepackage{units}
\usepackage{bbm}
\usepackage{booktabs}
\usepackage{array}
\usepackage{placeins}
\usepackage{epsfig}
\usepackage{lscape}
\usepackage{changes}
\usepackage{braket}
\usepackage{float}

\definecolor{myorange}{rgb}{1.0,0.27,0.0}

\bibstyle{apsrev.bib}

\begin{document}

\title{Spatial Organization of the Cytoskeleton enhances Cargo Delivery to Specific Target Areas 
on the Plasma Membrane of Spherical Cells}

\author{Anne E. Hafner}
\email{anne.hafner@lusi.uni-sb.de}
\author{Heiko Rieger}
\email{h.rieger@mx.uni-saarland.de}
\affiliation{Department of Theoretical Physics, Saarland University, 66123 Saarbr\"ucken, Germany}
\date{\today}


\begin{abstract}
\noindent
Intracellular transport is vital for the proper functioning and survival of a cell. Cargo
(proteins, vesicles, organelles, etc.) is transferred from its place of creation to its target
locations via molecular motor assisted transport along cytoskeletal filaments. The transport
efficiency is strongly affected by the spatial organization of the cytoskeleton, which
constitutes an inhomogeneous, complex network. In cells with a centrosome microtubules
grow radially from the central microtubule organizing center towards the cell periphery
whereas actin filaments form a dense meshwork, the actin cortex, underneath the cell
membrane with a broad range of orientations. The emerging ballistic motion along
filaments is frequently interrupted due to constricting intersection nodes or cycles of
detachment and reattachment processes in the crowded cytoplasm. In order to investigate
the efficiency of search strategies established by the cell's specific spatial organization of the
cytoskeleton we formulate a random velocity model with intermittent arrest states. With
extensive computer simulations we analyze the dependence of the mean first passage times
for narrow escape problems on the structural characteristics of the cytoskeleton, the motor
properties and the fraction of time spent in each state. We find that an inhomogeneous architecture with a small width of the actin cortex constitutes an efficient intracellular search strategy.
\end{abstract}


\maketitle

\section*{Introduction}
\label{sec:introduction}

\noindent
The accurate delivery of various cargoes is of great importance for maintaining the correct function of cells and organisms. Particles, such as vesicles, proteins, organelles, have to be transported to their specific destinations. In order to enable this cargo transfer, cells are equipped with a complex filament network and specialized motor proteins. The cytoskeleton serves as tracks for molecular motors. They convert the energy provided by ATP (adenosine triphosphate) hydrolysis into active motion along the cytoskeletal filaments, while they simultaneously bind to cargo \cite{Vale2000,Schliwa2003}. In addition to intracellular transport, the dynamic cytoskeleton and its associated motors also stabilize the cell shape, adjust it to different environmental circumstances, and drive cell motility or division \cite{TheCell}.\\

\noindent
The two main constituents of the cytoskeleton involved in intracellular transport are the polarized microtubules and actin filaments. In cells with a centrosome, the rigid microtubules grow radially from the central MTOC (microtubule organizing center) towards the cell periphery. In conjunction with the associated motor proteins kinesin and dynein, microtubules manage fast long-range transport between the cell center and periphery. In contrast to microtubules, which spread through the whole cell, actin filaments are mostly accumulated in a random fashion underneath the plasma membrane and construct the so called actin cortex \cite{TheCell}. Myosin motors operate on actin filaments and are therefore specialized for lateral transport in the cell periphery. Consequently, the cytoskeletal structure is very inhomogeneous and characterized by a 
thin actin cortical layer \cite{Salbreux2012,Eghiaian2015}.\\

\noindent
The `saltatory' transport \cite{Rebhun1967} by molecular motors is a cooperative mechanism. Several motors of diverse species are simultaneously attached to one cargo \cite{Welte2004,Vershinin2007,Balint2013,Hancock2014}. This enables a frequent exchange between actin and microtubule based transport, which is necessary for specific search problems. A prominent example of collaborative transport on actin and microtubule networks is the motion of pigment granules in fish and frog melanophores \cite{Rodionov2003, Gross2002}.
The activity level of the particular motor species, and thus the share in cooperation, is regulated by cell signaling \cite{Welte2004,Mallik2004}. 
In fish and frog, pigment granules are accumulated near the nucleus by extracellular stimuli transduced via PKA (protein kinase A) \cite{Rodionov2003, Gross2002}. The motor activity is thereby controlled via the level of cAMP (cyclic adenosine monophosphate). While low concentrations promote the action of dyneins, intermediate values stimulate myosin motors and high amounts of cAMP activate kinesins \cite{Rodionov2003}. Moreover, the infection of cells by adenoviruses triggers signaling through PKA and MAP (mitogen activated protein kinase), which enhances transport to the nucleus \cite{Suomalainen2001}; and the net transport of lipid droplets in Drosophila is directed to the cell center (periphery) by absence (presence) of the transacting factor Halo \cite{Gross2003,Welte2004}.\\

\noindent
Another aspect of intracellular transport is its intermittent nature. 
Molecular motors perform two phases of motility. Periods of directed active motion along cytoskeletal filaments interfere with effectively stationary states \cite{Bressloff2013}. Intersection nodes of the cytoskeleton cause the molecular motors to pause until they either manage to squeeze through the constriction and pass it on the same filament or switch to the crossing track and thus change the direction of the transported cargo \cite{Balint2013,Ali2007,Ross2008,Ross2008B}.
Motor proteins also detach of the filaments out of chemical reasons. In the cytoplasm the cargoes experience subdiffusive dynamics due to crowding effects \cite{Weiss2004}. The displacement is limited to the vicinity of the detachment site and negligible compared to the one of the active motion phase. Hence, detachment and reattachment processes effectively contribute to waiting times. However, cargo particles preferentially change their direction of motion at cytoskeletal intersections, which constitute motion barriers. The mean distance between two intersections, the mesh size of a network, is typically smaller than the processive run length of a single molecular motor \cite{Snider2004}.\\

\noindent
The spatial organization of the cytoskeleton as well as the activity of the different motor species and their behavior at network intersections establish a typical stochastic movement pattern of intracellular cargo, which suggests a random walk description \cite{Benichou2011,Bressloff2013}.
The narrow escape problem describes a common search process, where the target destination is represented by a specific, small region on the plasma membrane of a cell. Typical examples involve secretion processes in Cytotoxic T Lymphocytes (CTL) which play a key role in immune response and defeat tumorigenic or virus-infected cells. When in contact with an antigen-presenting cell, CTL form a connection with a diameter of the order of microns to it, which is called immunological synapse. Lytic granules are actively transported and released at the immunological synapse. They contain perforin and granzymes which induce apoptosis of the pathological cell. In order to prevent unintended damage of neighboring cells, the release of lytic granules is strictly confined to the immunological synapse. Thus, the transport of lytic granules towards the immunological synapse constitutes a narrow escape problem to be solved by CTL \cite{Grakoui1999,Qu2011,Angus2013,Ritter2013}. Moreover, the outgrowth of dendrites or axons from neurons \cite{Alberts2003,Chada2003} as well as repair mechanisms for corruptions of a cell's plasma membrane \cite{McNeil2005,Andrews2014} require the target-oriented transport of mitochondria and vesicles.\\

\noindent
In prior work on the narrow escape problem, Schuss et al. analytically investigated first passage properties of a purely diffusive searcher \cite{Schuss2007,Schuss2012}, while B\' enichou et al. identified benefits of search efficiency to small targets on the surface of spherical domains by intermittent phases of surface-mediated and bulk diffusion \cite{Benichou2010}.
Ballistic motion along the cytoskeleton and the different characteristics of microtubule- and actin-based transport was taken into account by Slepchenko et al. in order to model the spreading of pigment granules in melanophores \cite{Slepchenko2007}. They were able to determine the switching rate between microtubules and actin filaments by fitting their theoretical results to experimental data of aggregation and dispersion of pigment granules in fish melanophores, whose cytoskeleton appears to be rather homogeneous. 
In order to study first passage properties of cargo transport from the nucleus to the complete plasma membrane, Ando et al. recently considered an inhomogeneous network distribution in which the cytoskeleton is confined to a delimited shell \cite{Ando2015}. Within a continuum model of increased bulk diffusion, they found that the transit time can be significantly reduced when the cytoskeletal shell is placed close to the nucleus. Moreover, they explicitly modeled cytoskeletal networks and investigated the impact of number, length and polarity of filaments as well as detachment and reattachment processes by considering intermittent phases of ballistic transport and cytoplasmic diffusion. Via extensive computer simulations, they showed, inter alia, that an outward-directed network polarity expectedly improves transit times while the actual distribution of filament orientations is less effective. 
Nonetheless, transport in cells comprises an intricate interplay between motor performance and spatial organization of the cytoskeleton, which is generally inhomogeneous itself. The description of this interaction is a challenging theoretical task and we still lack precise knowledge about how cells adapt to various transport tasks and especially to narrow escape problems.\\

\noindent
In the following, we present a coarse grained model of intracellular transport by considering the effective movement between network nodes, while discarding the single steps of individual motors at the molecular level. We introduce a random velocity model with intermittent arrest states where the dynamic cytoskeleton is implicitly modeled by probability density functions for network orientation and mesh size. 
The proposed model allows the study of diverse transport tasks. Here we focus on the narrow escape problem and address the effects of the interplay between inhomogeneous cytoskeletal architecture and motor performance on the search efficiency to small targets alongside the membrane of spherical cells.

\section*{Model}
\label{sec:model}

\subsection*{General Random Velocity Model}

\begin{figure*}[t]
\centering
\includegraphics[width=\linewidth]{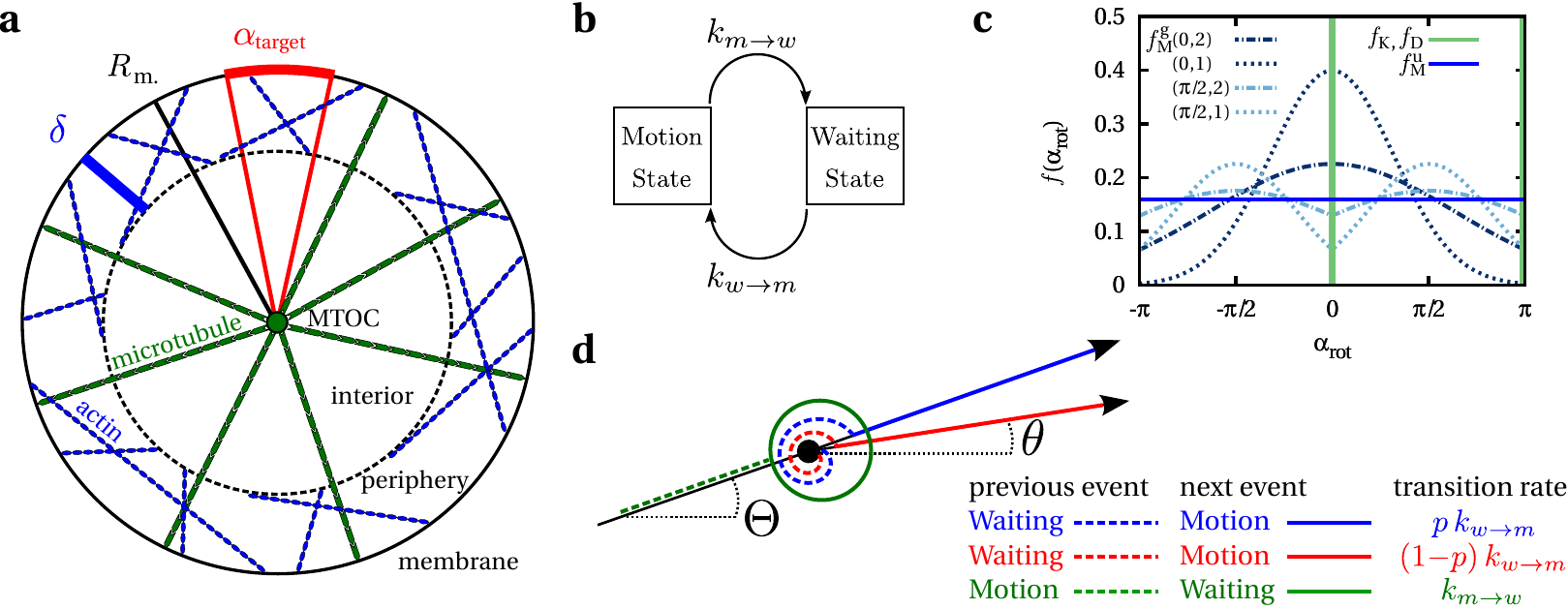}
\caption
{Random Velocity Model for Intracellular Search.
\textbf{a} The model geometry displays the confined and inhomogeneous architecture of a cell of radius $R_\text{membrane}$. While the interior is filled with radial microtubules, the periphery is dominated by random actin filaments. The width of the actin cortical layer is denoted by $\delta$ and the narrow escape hole has an angular diameter $\alpha_\text{target}$. \textbf{b} Pausing processes arise due to constricting intersection nodes of the cytoskeletal network. This is accounted in the model by two motility states with specific transition rates $k_{m\rightarrow w}$ and $k_{w\rightarrow m}$. \textbf{c} The rotation angle distributions $f_\text{K}$ and $f_\text{D}$ for kinesins and dyneins acting on microtubules are delta-peaked, while the ones of myosins may be uniformly distributed ($f^u_\text{M}$) or gaussian distributed ($f^g_\text{M}(\mu,\sigma)$) with various expectation values $\mu$ and standard deviations $\sigma$ regarding the orientation of actin filaments. \textbf{d} Sketch of the walk during two consecutive events and the corresponding transition rates, where $\theta$ denotes the new and $\Theta$ the previous direction.
}
\label{figure1}
\end{figure*}

\noindent
Cells establish specific search strategies for cargo transport by alterations of the cytoskeletal organization and regulation of the motor behavior at network intersections. In order to study the efficiency of spatially inhomogeneous search strategies we formulate a random velocity model in continuous space and time composed of two states of motility: (i) a ballistic motion state at constant speed $v{=}1$, which corresponds to active transport by molecular motors in between two successive intersections of the filamentous network and (ii) a waiting state, which is associated to pauses at intersection nodes of the cytoskeleton. The swaps from one state to another are arranged via constant but generally asymmetric transition rates $k_{m\rightarrow w}$ for a switch from motion to waiting and $k_{w\rightarrow m}$ for an inverse transition, see figure \ref{figure1} (b). These lead to exponentially distributed time periods $t_m$, $t_w$ spent in each state of motility
\begin{align}
p(t_m) &= k_{m\rightarrow w} e^{-k_{m\rightarrow w}t_m}, \\
p(t_w) &= k_{w\rightarrow m} e^{-k_{w\rightarrow m}t_w},
\end{align}
and mean residence times of $1/k_{m\rightarrow w}$, $1/k_{w\rightarrow m}$ for the motion and waiting state, respectively, which is biologically consistent as active lifetimes of cargo particles are exponentially distributed \cite{Arcizet2008}. The event rates are directly connected to biologically tractable properties of the cytoskeleton and the motor proteins. The mesh size $\ell$ of the underlying cytoskeletal network, which reflects the typical distance between two consecutive intersections, defines the rate $k_{m\rightarrow w} {=} v/\ell$ for a transition from the ballistic motion to the pausing state. Whereas the event rate $k_{w\rightarrow m}$ is determined by the characteristic waiting time at an intersection node.\\

\noindent
Whenever a particle has reached a network intersection and paused, it may either keep moving processively along the same filament with probability $p$ or it may change to a crossing track with probability $(1{-}p)$, as sketched in figure \ref{figure1} (d). This provides a typical timescale $[(1-p)k_{w\rightarrow m}]^{-1}$ ($[pk_{w\rightarrow m}]^{-1}$) of changing (remaining on) the filament subsequent to a waiting period. With regard to the rotational symmetry of a cell, the new direction $\theta {=} \phi +\alpha_\text{rot}$ is always chosen with respect to the radial direction $\tan(\phi){=}y/x$. The rotation angle $\alpha_\text{rot}$ is drawn from a distribution $f(\alpha_\text{rot})$ which is characteristic for the underlying cytoskeletal network. Due to the inhomogeneous structure of the cytoskeleton, $f(\alpha_\text{rot})$ depends on the location of the particle inside the cell.

\subsection*{Model Geometry}

\noindent
Our model system is designed according to the inhomogeneous internal organization of a cell, see figure \ref{figure1} (a). We assume a circular confined geometry of radius $R_\text{membrane}$, which displays the plasma membrane and will be fixed in the following ($R_\text{membrane}{=}1$). The cytoplasm is split into an interior region, where only microtubules are present, and a periphery, which is dominated by the actin cortex but may also be pervaded by microtubules. The width of the actin cortical layer is denoted by $\delta$, so that an internal margin of radius $R_\text{internal}{=}R_\text{membrane}-\delta$ emerges. Throughout this article, the target destination of the cargo is assumed to be a narrow escape hole in the plasma membrane with opening angle $\alpha_\text{target}$.

\subsection*{Random Velocity Model in the Interior}

\noindent
The interior of a cell is controlled by the radial network of microtubules with its associated kinesin and dynein motors. Since they manage fast long-range transport inside a cell, the rate to switch from the motion to the waiting state $k^i_{m\rightarrow w} {=} 0$ is fixed to zero for simplicity. This leads to uninterrupted radial movement along the internal microtubule network. However, the particle may be forced into the waiting state due to confinement events, as dyneins are assumed to stop at the central MTOC. Hence the transition rate to the motion state is set to a non-zero value $k^i_{w\rightarrow m}$. 

\subsection*{Random Velocity Model in the Periphery}

\noindent
Due to the complex network structure of the periphery the searcher frequently encounters intersection nodes at rate $k^p_{m\rightarrow w}$ and switches to the waiting state. Subsequently, the particle may either keep moving along the previously used track at rate $p \, k^p_{w\rightarrow m}$ or it may change to a crossing filament at rate $(1-p) \, k^p_{w\rightarrow m}$. The rotation angle is thereby drawn of a distribution
\begin{align}
f(\alpha_\text{rot}) = q_\text{K} \, f_\text{K}(\alpha_\text{rot})+q_\text{D} \, f_\text{D}(\alpha_\text{rot})+q_\text{M} \, f_\text{M}(\alpha_\text{rot}),
\end{align}
which specifies the peripheral environment in terms of the filament orientation ($f_i(\alpha_\text{rot})$) and motor species activity ($q_i$). The probabilities $q_\text{K}$, $q_\text{D}$, $q_\text{M}$ correspond to directional changes induced by kinesin, dynein and myosin, respectively, with $q_\text{K}+q_\text{D}+q_\text{M}=1$. With regard to the radial orientation of microtubules and the directionality of the motors, the rotation angle distributions associated to kinesins and dyneins are delta peaked
\begin{align}
f_\text{K}(\alpha_\text{rot}) & = \delta(\alpha_\text{rot}),\quad~~\,\qquad\quad\text{ for } \alpha_\text{rot} \in ({-}\pi;\pi],\\
f_\text{D}(\alpha_\text{rot}) & = \delta(\alpha_\text{rot}-\pi),\qquad\quad\text{ for } \alpha_\text{rot} \in ({-}\pi;\pi].
\end{align}
In case of a directional change initiated by myosins, the rotation angle distribution is assumed to be either uniform or cut-off-gaussian, which takes into account the randomness of actin networks
\begin{align}
f_\text{M}^\text{u}(\alpha_\text{rot}) & = \frac{1}{2\pi},\qquad\qquad\qquad\quad\text{for } \alpha_\text{rot} \in ({-}\pi;\pi],\\
f_\text{M}^\text{g}(\alpha_\text{rot}) & = \frac{1}{2} f^+ + \frac{1}{2} f^-,
\label{eq:rotanglemyosin}
\end{align}
with
\begin{align}
f^+ & = \frac{\mathcal{N}}{\sigma\sqrt{2\pi}}\exp\left({-}\frac{(\alpha_\text{rot}{-}\mu)^2}{2\sigma^2}\right),\text{ for } \alpha_\text{rot} \in [0;\pi],\\
f^- & = \frac{\mathcal{N}}{\sigma\sqrt{2\pi}}\exp\left({-}\frac{(\alpha_\text{rot}{+}\mu)^2}{2\sigma^2}\right),\text{ for } \alpha_\text{rot} \in [{-}\pi;0],
\end{align}
where $\mu>0$ and $\mathcal{N}$ denotes the normalization constant, see figure \ref{figure1} (c). In conclusion, the peripheral network is characterized by a mean mesh size $1/k_{m\rightarrow w}$, its structure is reflected by the rotation angle distributions $f_i(\alpha_\text{rot})$ and the motor activity is defined via $q_i$.

\begin{figure*}[t]
\centering
\includegraphics[width=\linewidth]{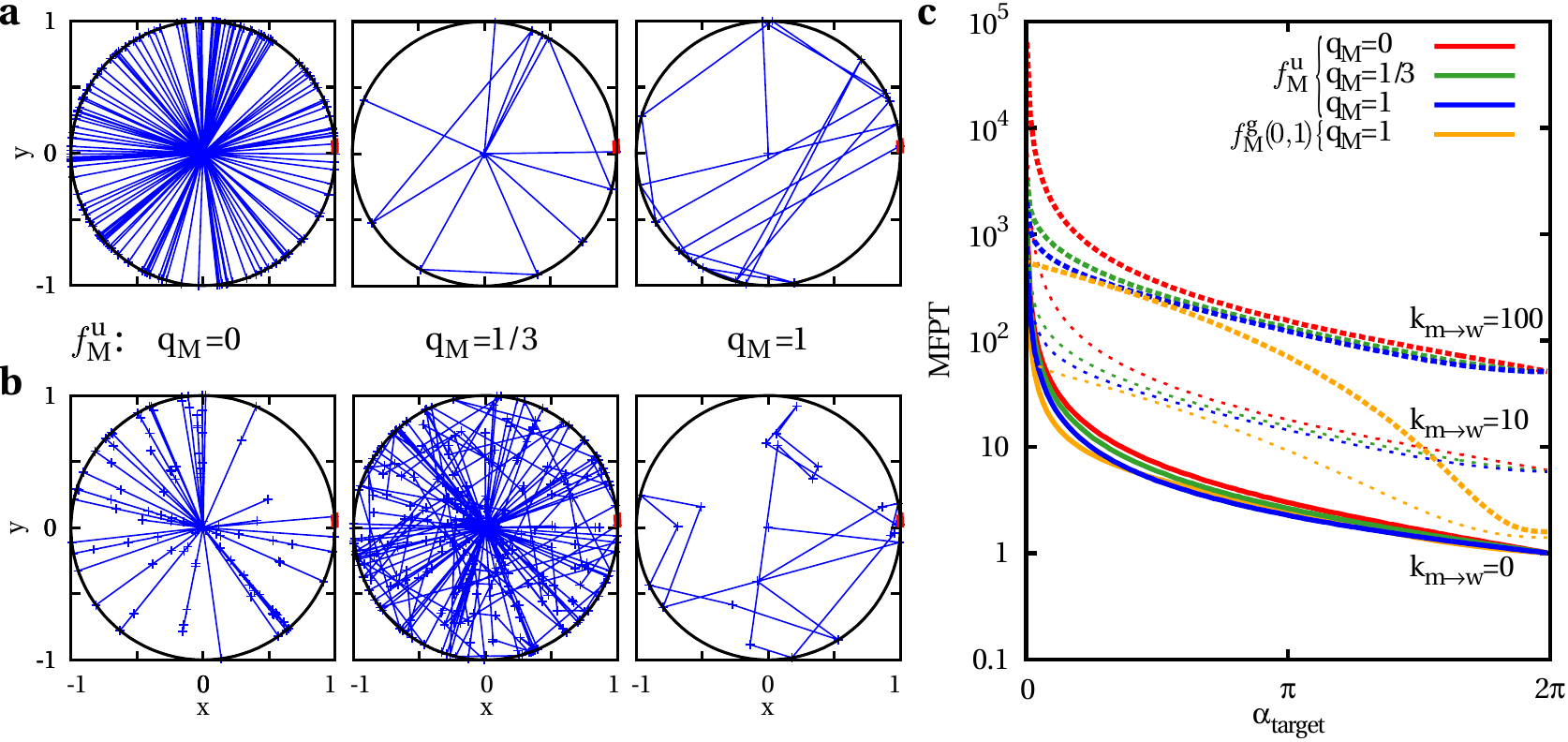}
\caption
{Homogeneous Search Strategy. \textbf{a} Sample trajectories (blue lines) during homogeneous search strategies on a circle of radius $R_\text{membrane}{=}1$ with a membranous narrow escape of opening angle $\alpha_\text{target}{=}0.1$ (red zone). The transition rate to the waiting state is fixed to $k_{m\rightarrow w}{=}0$. Consequently, the cargo particle does not encounter any network intersections and hence does not change its direction of motion inside the bulk. For $q_\text{M}{=}0$ the network is completely radial, whereas it is uniformly distributed in the case of $q_\text{M}{=}1$. Intermediate values of the probability $q_\text{M}$ correspond to homogeneous combinations of radial microtubules and random actin filaments, which is reflected by the trajectories. \textbf{b} Sample trajectories for $k_{m\rightarrow w}{=}1$. The searcher frequently changes its direction at the network intersections (blue crosses) according to the probabilities $q_\text{M}$, $q_\text{K}{=}q_\text{D}{=}(1-q_\text{M})/2$ and their associated rotation angle distributions until it reaches the membranous target zone $\alpha_\text{target}{=}0{.}1$. \textbf{c} MFPT in dependence of the target size $\alpha_\text{target}$ for different values of the transition rate $k_{m\rightarrow w}$ and various motor activities $q_\text{M}$, $q_\text{K}{=}q_\text{D}{=}(1-q_\text{M})/2$ for uniformly distributed ($f_\text{M}^\text{u}$) as well as outward-directed ($f_\text{M}^\text{g}(\mu{=}0,\sigma{=}1)$) actin structures.}
\label{figure2}
\end{figure*}

\begin{figure*}[t]
\centering
\includegraphics[width=\linewidth]{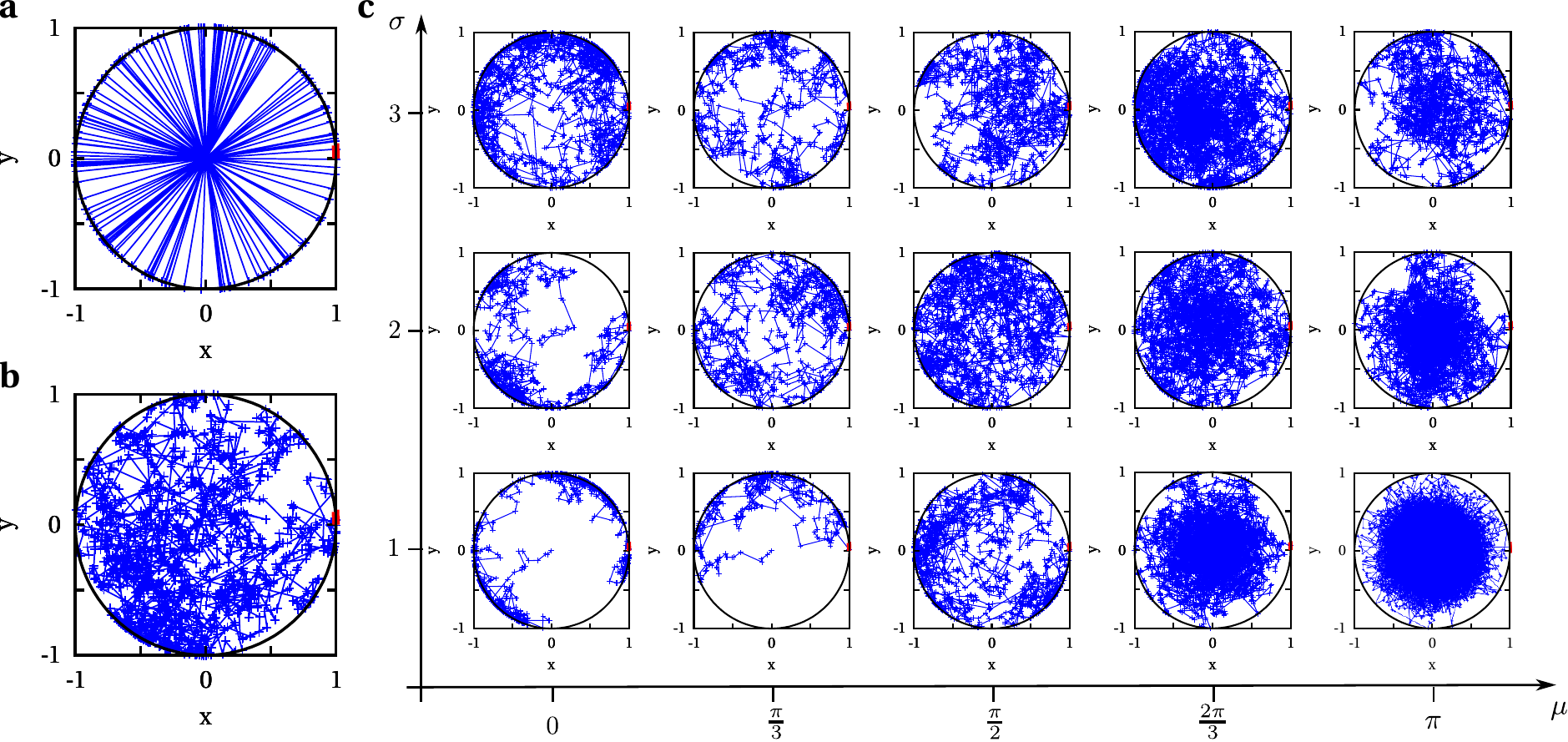}
\caption
{Sample Trajectories for Homogeneous Search Strategies with $\alpha_\text{target}{=}0{.}1$. \textbf{a} Motion on a radial microtubule network ($q_\text{M}{=}0$), where $k_{m\rightarrow w}{=}0$. \textbf{b} Trajectory for a cargo moving along a uniformly random actin network ($q_\text{M}{=}1$, $f_\text{M}^\text{u}$) with $k_{m\rightarrow w}{=}10$. \textbf{c} Motion on a gaussian distributed actin network ($q_\text{M}{=}1$, $f_\text{M}^\text{g}(\mu,\sigma)$) with a mesh size defined by $k_{m\rightarrow w}{=}10$ for various $\mu$ and $\sigma$. For instance $\sigma{=}1$ demonstrates the diverse residence areas of the cargo with respect to $\mu$. While for $\mu<\pi/2$ the particle predominantly moves close to the boundary, for $\mu > \pi/2$ it is pushed away from the membrane. By increasing $\sigma$ the motion pattern gets randomized due to the broadening of the rotation angle distribution $f_\text{M}^\text{g}(\mu,\sigma)$.
}
\label{figure2b}
\end{figure*}

\subsection*{Confinement Events}

\noindent
The spatial geometry of the model system imposes various confinement events, which are further specified in the following.
At the onset, each searcher is assumed to start its walk in the center of the cell, where it is linked to a kinesin and runs in a uniformly distributed initial direction.
As soon as a particle encounters the outer membrane margin at radius $R_\text{membrane}$ it will switch into the pausing state, since it is assumed to detach of the filament, and check for the target zone $\alpha_\text{target}$. If it is found, the walk will be terminated, otherwise the particle will wait at rate $(1-p) \, k^p_{w\rightarrow m}$ and the rotation angle distribution will be restricted to allowed values.
The same holds in the case that a cargo transported by myosin hits the internal margin $R_\text{internal}$, which is created by the structural inhomogeneity of the cytoskeleton. Crossovers of the internal margin by kinesins or dyneins, happen uninterruptedly under a change of the characteristic event rates for interior and periphery.
Whenever a dynein coupled particle reaches the MTOC in the center of the cell, it will wait with rate $k^i_{w\rightarrow m}$ before it will change to kinesin motion in a uniformly distributed direction.
Consequently, the mean waiting time at confinement events is not necessarily the same as the one at network intersections in the bulk. In general it will be larger.

\section*{Results}

\noindent
In the following, we perform extensive Monte Carlo simulations in order to analyze the dependence of the search efficiency to narrow escapes alongside a cell's membrane on the spatial organization of the cytoskeleton as well as the motor performance at network intersections. For that purpose we define the mean first passage time (MFPT) to a target as the ensemble average over first passage events of $5\times 10^5$ independent realizations of the walk, in which all cargoes are initially located at the center of the cell and start moving in an uniformly distributed direction $\Theta\in[-\pi;\pi)$. At first, we will consider search strategies, where the motors operate on homogeneous cytoskeletal networks. Then, we will focus on the inhomogeneous architecture and elaborate the influence of the actin cortical width on the mean first passage properties of narrow escape problems.

\begin{figure*}[t]
\centering
\includegraphics[width=0.97\linewidth]{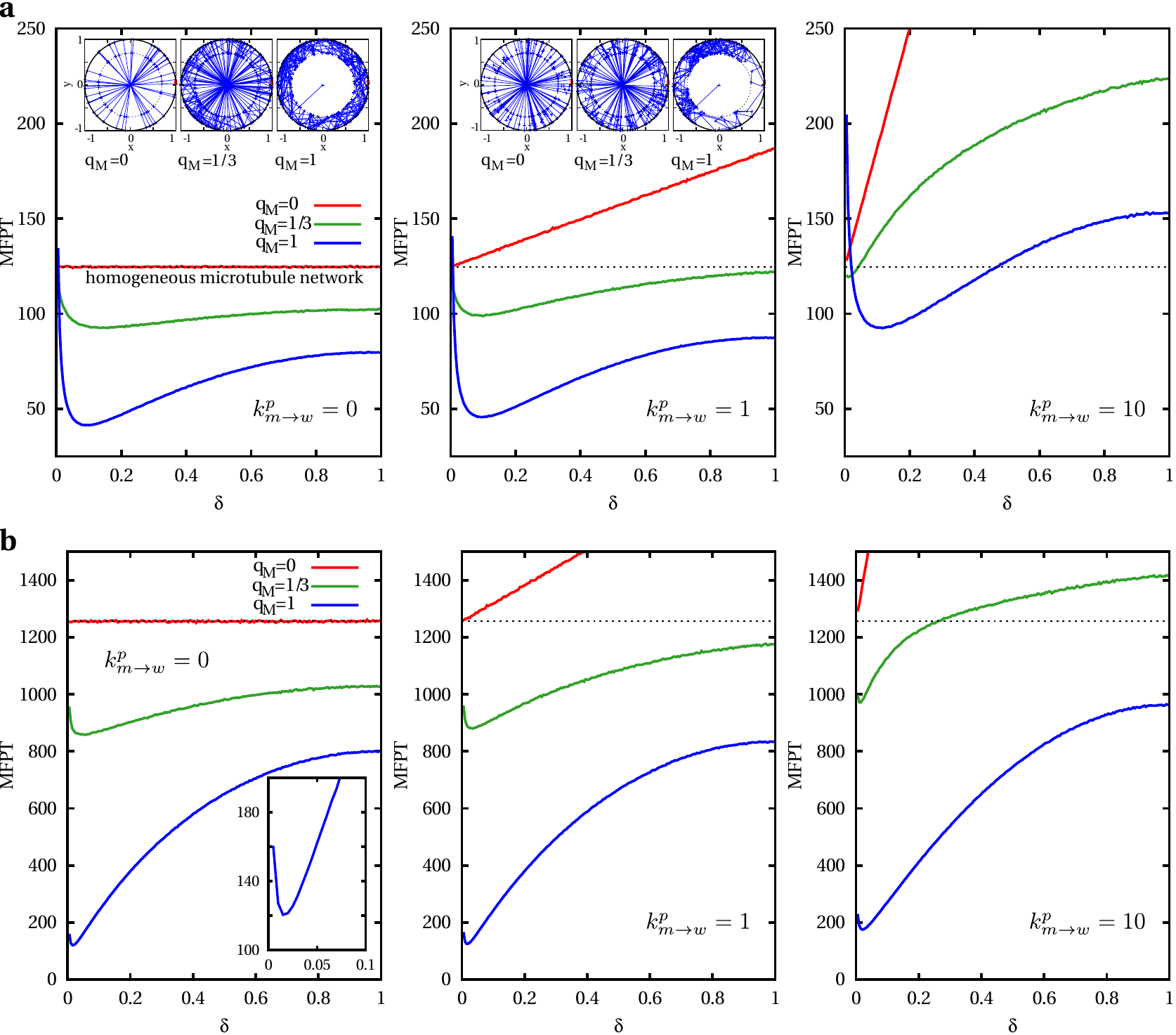}
\caption
{Inhomogeneous Search Strategy. 
Interior parameters: $k^i_{m\rightarrow w}{=}0$, $k^i_{w\rightarrow m}{=}\infty$; periphery parameters: $k^p_{m\rightarrow w}\in \{0;1;10\}$, $k^p_{w\rightarrow m}{=}\infty$, $p{=}0$, $q_\text{K}{=}q_\text{D}{=}(1-q_\text{M})/2$, $q_\text{M}\in \{0;1/3;1\}$; target size: $\alpha_\text{target}\in\{0{.}1;0{.}01\}$. \textbf{a} MFPT to a narrow escape of opening angle $\alpha_\text{target}{=}0{.}1$ for an inhomogeneous cytoskeleton in dependence of the actin cortical width $\delta$. Since $k^p_{w\rightarrow m}{=}\infty$, the searcher changes its direction instantaneously at each cytoskeletal intersection in the periphery, which occur at various rates $k^p_{m\rightarrow w}$. The new direction is thereby chosen with probability $q_\text{M}$ according to a uniformly rotation angle distribution $f_\text{M}^\text{u}(\alpha_\text{rot})$, or with probabilities $q_\text{K}{=}q_\text{D}{=}(1-q_\text{M})/2$ according to a rotation angle drawn from $f_\text{K}(\alpha_\text{rot})$, $f_\text{D}(\alpha_\text{rot})$, respectively, thus the motors are not processive $p{=}0$ at network intersections. The dashed black line refers to a homogeneous search strategy on a pure and radial microtubule network. The insets show some sample trajectories (blue lines) for $\delta{=}0.3$ and different values of $q_\text{M}$. \textbf{b} MFPT to a narrow opening of angular diameter $\alpha_\text{target}{=}0{.}01$. The inset is a detail of $q_\text{M}{=}1$ for $\delta \in \left[0;0{.}1\right]$, which emphasized the shift of the optimal cortical width by a decrease of the target size.
}
\label{figure4}
\end{figure*}

\subsection*{Homogeneous Search Strategy}

\noindent
Here, we neglect the inhomogeneous structure of the cytoskeleton and assume that it spreads through the whole cell in a homogeneous manner. Since we aim to isolate the impact of the cytoskeletal architecture, we ignore the motor's processivity ($p{=}0$) and waiting processes ($k_{w\rightarrow m}{=}\infty$). Hence at each network intersection the particle immediately changes its direction according to
\begin{align}
f(\alpha_\text{rot}) = q_\text{K} \, f_\text{K}(\alpha_\text{rot})+q_\text{D} \, f_\text{D}(\alpha_\text{rot})+q_\text{M} \, f_\text{M}(\alpha_\text{rot}),
\end{align}
where $f_\text{M}\in\{f_\text{M}^\text{u};f_\text{M}^\text{g}(\mu,\sigma)\}$ and $q_\text{K}=q_\text{D}=(1-q_\text{M})/2$. In the case of $q_\text{M}{=}0$ myosin motors are deactivated and the transport is managed by kinesin and dynein on a radial network of microtubules, while $q_\text{M}{=}1$ leads to a pure actin mesh with myosins and either uniformly or gaussian distributed filaments. Intermediate values of $q_\text{M}$ correspond to cooperative transport on homogeneous combinations of microtubules and actin filaments with active kinesins, dyneins and myosins. Sample trajectories which implicitly reflect the network structure are given in figure \ref{figure2} for uniformly random actin orientations.\\

\noindent
In order to investigate the influence of the network composition and motor activity, we measure the MFPT in dependence of the target size $\alpha_\text{target}$ for different event rates $k_{m\rightarrow w}$ and various probabilities $q_\text{M}$. As expected, the MFPT increases monotonically with decreasing opening angle $\alpha_\text{target}$, see figure \ref{figure2} (c). Apparently, directional changes at intersections of a homogeneous cytoskeletal network do not improve the search efficiency nor does a cooperative behavior of different motor species ($q_\text{M}\in (0;1)$). The MFPT increases with greater transition rate $k_{m\rightarrow w}$ and is quite robust against alterations of the probability $q_\text{M}$. However, we notice that a random actin mesh ($q_\text{M}{=}1$) is preferable to a radial microtubule network ($q_\text{M}{=}0$).\\

\noindent
As expected, figure \ref{figure2} (c) shows that an outward-oriented actin network with $f_\text{M}^\text{g}(\mu,\sigma)$ and for example $\mu{=}0\!<\!\pi/2$, $\sigma{=}1$ is advantageous for large target sizes. This was also stated by Ando et al. for transport from the cell center to the whole membrane, i.e. $\alpha_\text{target}{=}2\pi$ \cite{Ando2015}. Remarkably, the search for small targets does also benefit from outward-directed actin filaments. While inward-directed ($\mu>\pi/2$) actin polarities drive the cargo towards the cell center, which lowers the probability to reach the membrane and detect the target, outward-directed ($\mu<\pi/2$) actin networks push the particle towards the plasma membrane, as indicated by the sample trajectories in figure \ref{figure2b}. This introduces a topologically induced technique to scan the membrane for small targets as the cargo is predominantly moving in vicinity of the cell boundary. This effect vanishes for increasing $\sigma$, since the resulting broadening of the distribution $f_\text{M}^\text{g}(\mu,\sigma)$ leads to approximately uniformly distributed filaments ($f_\text{M}^\text{u}$) and thus a randomization of the search. Consequently, as evident from figure \ref{figure2} (c) for small target sizes, a homogeneous outward-directed actin structure may be more efficient (even for $k_{m\rightarrow w}\neq 0$) than the best search strategy on a homogeneous microtubule network (i.e. $k_{m\rightarrow w}{=}0$). The reason is that large excursions to the cell center are inhibited by outward-directed actin polarities, while detours to the center are necessary for transport along microtubules in order to change radial direction and reach the target.

\subsection*{Inhomogeneous Search Strategy}

\noindent
In case of a homogeneous cytoskeleton, the most efficient search strategy is an uninterrupted motion on a pure actin network without any directional changes in the bulk. Such a motion scheme is sufficient for large membranous targets, but generally fails for narrow escape problems. May an inhomogeneous search strategy, like it is found in living cells, be the key to the efficient detection of small departure zones on the plasma membrane? Guided by this question, we check for the influence of the actin cortical width $\delta$ on the effectiveness of the narrow escape search problem.

\subsubsection*{Influence of the Actin Cortical Width}

\noindent
In order to investigate the pristine effect of an inhomogeneous cytoskeleton, we perform Monte Carlo simulations and assume $k^i_{m\rightarrow w}{=}0$, $k^i_{w\rightarrow m}{=}\infty$ in the interior and $k^p_{m\rightarrow w}\in \{0;1;10\}$, $k^p_{w\rightarrow m}{=}\infty$, $p{=}0$ in the periphery. Consequently, the cargo changes its direction of motion instantaneously at each intersection node. Figure \ref{figure4} displays the resulting MFPT to small membranous target sites ($\alpha_\text{target}{=}0{.}1$ and $\alpha_\text{target}{=}0{.}01$) in dependence of the actin cortical width $\delta$ for different peripheral mesh sizes defined by the node encountering rate $k^p_{m\rightarrow w}$ and motor activities $q_\text{M}$, $q_\text{K}{=}q_\text{D}{=}(1-q_\text{M})/2$ for $f_\text{M}^\text{u}$.\\

\noindent
Remarkably, we find that the MFPT responds sensitively to alterations of the motor activity $q_\text{M}$ and exhibits a nontrivial minimum for $q_\text{M}\neq 0$ and small targets. In the case of $q_\text{M}{=}0$, transport is solely managed by kinesins and dyneins on a radial but generally inhomogeneous microtubule network. The inhomogeneity is determined by $\delta$ and results in a change of the mesh size from the cell's interior (with $k^i_{m\rightarrow w}{=}0$) to the periphery (with various $k^p_{m\rightarrow w}$). Hence, for $k^p_{m\rightarrow w}{=}k^i_{m\rightarrow w}{=}0$ the network is homogeneous and the MFPT is constant in $\delta$, as shown in figure \ref{figure4}. Contrarily, the particle frequently changes its direction in the peripheral bulk according to $k^p_{m\rightarrow w}\neq 0$, which leads to a local back-and-forth motion of the cargo on the same filament for $q_\text{M}{=}0$. This back-and-forth pattern significantly hinders both the hitting with the membrane as well as the retraction to the central MTOC, which is necessary in order to change radial direction and detect the membranous target. Consequently, the MFPT monotonically increases with growing cortical width $\delta$, because a larger peripheral area increases the interruption by back-and-forth motion. In the case of $q_\text{M}\neq 0$, the minimum of the MFPT for $\delta \in \left(0;1\right)$ is most prominent for $q_\text{M}{=}1$, i.e. a peripheral motion dominated by transport along actin filaments, achieved by a high activity level of myosins, is most efficient. Contrarily, a high activity of dyneins would be favourable for virus trafficking or aggregation of pigment granules near the nucleus \cite{Slepchenko2007,Rodionov2003}. But for small membranous targets, inhomogeneous networks lead to a considerable gain of efficiency in comparison to the corresponding homogeneous limits $\delta\!\to\! 1$ and $\delta\!\to\! 0$. The case $\delta{=}1$ leads to a homogeneous search strategy which is defined by the features of the cell's periphery, i.e. the network possesses a mesh size defined by $k^p_{m\rightarrow w}$ and is composed of actin filaments and microtubules according to the probabilities $q_\text{M}$, $q_\text{K}{=}q_\text{D}{=}(1-q_\text{M})/2$. For instance $q_\text{M}{=}1$ and $\delta{=}1$ corresponds to intracellular transport on a homogeneous and uniformly random actin network. Contrarily, for $\delta{=}0$ the cargo's characteristics are purely determined by the cell's interior via $k^i_{m\rightarrow w}{=}0$, $k^i_{w\rightarrow m}{=}\infty$, which leads to motion along a homogeneous network of radial microtubules. \\

\noindent
While the MFPT is continuous in the limit $\delta\!\to\! 1$, it diverges for $\delta \!\to\! 0$ and $q_{M}\neq 0$. Consequently, a comparison to MFPT$(\delta{=}0)$ is not directly possible and we include this limit, which we studied in figure \ref{figure2}, by the black dashed line in figure \ref{figure4} and following figures. The divergence of the MFPT for $\delta\!\to\! 0$ originates from a motility restriction by a narrow cortex. Cargo particles which are transported by myosins get localized for narrow actin cortices, since the resulting prompt collisions with the two margins at radii $R_\text{internal}$ and $R_\text{membrane}$ inhibit substantially large displacements. Thus, the actin cortex can also act as a barrier for transport to a cell's plasma membrane.
The results shown in figure \ref{figure4} visualize that an increased chance of directional alterations by increase of $k^p_{m\rightarrow w}$, leads to a loss of search efficiency. This directly suggests a profit by a processive behavior of motor proteins at intersection nodes. Furthermore, a decrease in the opening angle $\alpha_\text{target}$ provokes a more prominent minimum of the MFPT at lower cortical widths $\delta$, compare figure \ref{figure4} (a) and (b).

\begin{figure*}[ht]
\centering
\includegraphics[width=\linewidth]{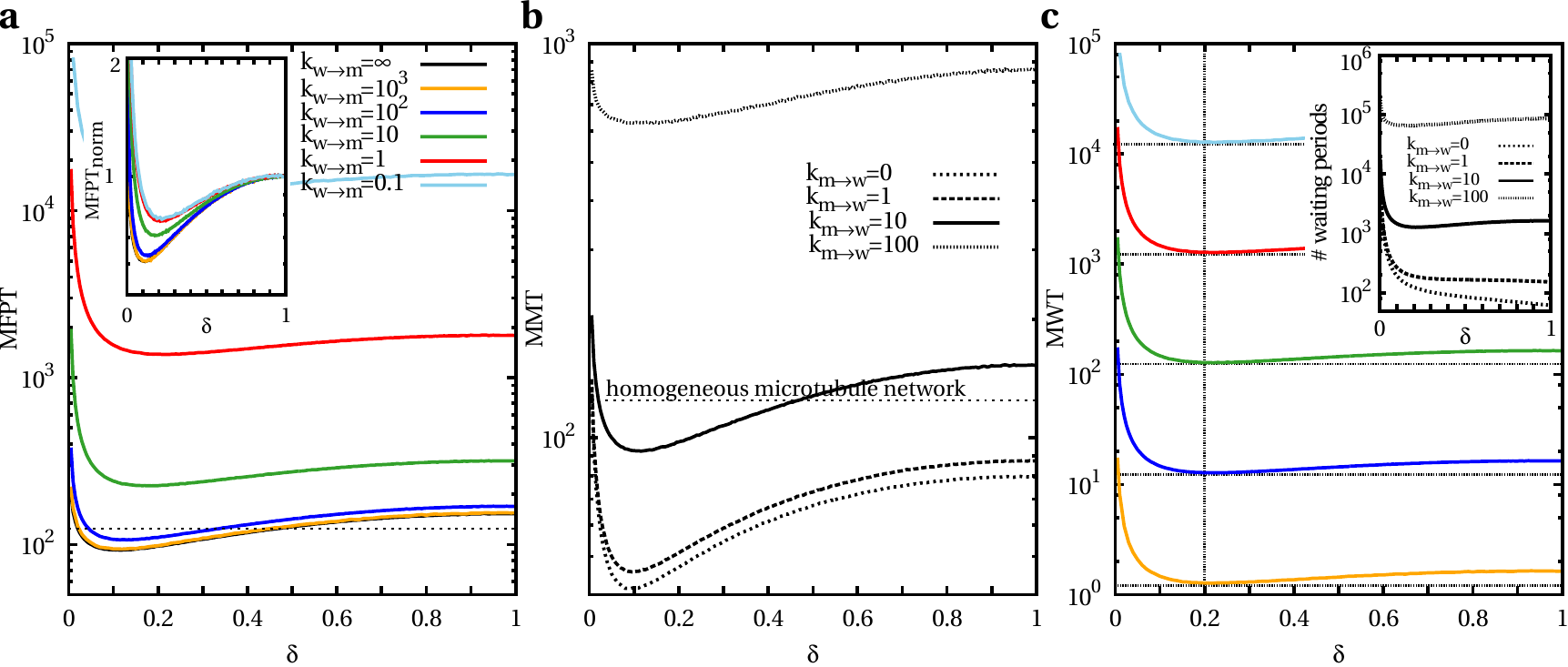}
\caption
{Influence of the Waiting Time Distribution on Inhomogeneous Search Strategies. 
Interior parameters: $k^i_{m\rightarrow w}{=}0$, $k^i_{w\rightarrow m}{=}k_{w\rightarrow m}$; periphery parameters: $k^p_{m\rightarrow w}{=}k_{m\rightarrow w}$, $k^p_{w\rightarrow m}{=}k_{w\rightarrow m}$, $p{=}q_\text{K}{=}q_\text{D}{=}0$, $q_\text{M}{=}1$; target size: $\alpha_\text{target}{=}0{.}1$. \textbf{a} MFPT for $k_{m\rightarrow w}{=}10$ and various $k_{w\rightarrow m}$ in dependence of the actin cortex width $\delta$. The case of $k_{w\rightarrow m}{=}\infty$ corresponds to instantaneous directional changes. A systematic increase in the mean waiting time per arrest state, i.e. a decrease in $k_{w\rightarrow m}$, heightens the MFPT. Moreover, the optimal value of the actin cortical width $\delta_\text{opt}$, which minimizes the MFPT, is shifted for decreasing transition rates $k_{w\rightarrow m}$. This is further illustrated by the inset, where the MFPT is normalized according to $\text{MFPT}_\text{norm}=\text{MFPT}/\text{MFPT}(\delta{=}1)$. \textbf{b} The MMT depends on the rate $k_{m\rightarrow w}$ and exhibits a minimum for $\delta$. It does not depend on the event rate $k_{w\rightarrow m}$ and is of course equivalent to the MFPT without any waiting processes ($k_{w\rightarrow m}{=}\infty$). \textbf{c} The MWT for $k_{m\rightarrow w}{=}10$ displays a minimum and is shifted by decreasing transition rates $k_{w\rightarrow m}$ (the colors correspond to part a). This is further illustrated by the inset, a transformation to the mean number of waiting periods via $\text{MWT}\!\times\! k_{w\rightarrow m}$ is applied for various $k_{m\rightarrow w}$.
}
\label{figure5}
\end{figure*}

\subsubsection*{Influence of the Waiting Time Distribution}

\noindent
So far the waiting times, which are induced by intersection nodes and confinement events, have been neglected by fixing the rate to switch from waiting to motion to infinity, $k^i_{w\rightarrow m}{=}\infty$, $k^p_{w\rightarrow m}{=}\infty$. Here we address the impact of the mean waiting time per arrest state. For that purpose we assume radial microtubules in the cell interior, $k^i_{m\rightarrow w}{=}0$, $k^i_{w\rightarrow m}{=}k_{w\rightarrow m}$. The motion in the periphery is non-processive ($p{=}0$) and performed on a random actin network ($q_\text{M}{=}1$) of a given mesh size, defined by $k^p_{m\rightarrow w}{=}k_{m\rightarrow w}$. The mean waiting time is determined by $k^p_{w\rightarrow m}{=}k_{w\rightarrow m}$ and the target size is fixed to $\alpha_\text{target}{=}0{.}1$.\\

\noindent
As expected, a systematic decrease of the rate $k_{w\rightarrow m}$, i.e. increase of the mean waiting time per arrest state, extends the MFPT in comparison to instantaneous directional changes for $k_{w\rightarrow m}{=}\infty$, as shown in figure \ref{figure5} (a). Remarkably, the position of the minimum, which determines the optimal cortical width $\delta_\text{opt}$, is shifted by a reduction in $k_{w\rightarrow m}$. This is further illustrated in the inset of figure \ref{figure5} (a) by normalization of the MFPT to $\text{MFPT}_\text{norm}=\text{MFPT}/\text{MFPT}(\delta{=}1)$.\\

\noindent
The change in $\delta_\text{opt}$ is based on the enhanced impact of waiting times on the first passage properties. The MFPT is composed of the total mean motion time (MMT) and the total mean waiting time (MWT) which the cargo experiences in the course of the search. figure \ref{figure5} (b) shows that the MMT displays pronounced minima and is indeed independent of the rate $k_{w\rightarrow m}$ and thus fully determined by the mesh size of the network given via $k_{m\rightarrow w}$. In contrast to that, the MWT, given in figure \ref{figure5} (c), depends on both rates $k_{m\rightarrow w}$ and $k_{w\rightarrow m}$. Since the rate $k_{w\rightarrow m}$ determines the mean waiting time per arrest state, it influences the MWT only via a multiplicative factor, as obvious by the shift in figure \ref{figure5} (c) for $k_{m\rightarrow w}{=}10$. The impact of the transition rate $k_{w\rightarrow m}$ on the MWT can be excluded by the transformation to the mean number of waiting periods
\begin{align}
\#\,\text{waiting periods} = \text{MWT} \times k_{w\rightarrow m}.
\end{align}
The inset of figure \ref{figure5} (c) displays its dependence on $k_{m\rightarrow w}$ and $\delta$. Remarkably, the number of waiting periods may also exhibit a minimum for small cortical widths.\\

\noindent
Consequently, we can solve the question of the shift in $\delta_\text{opt}$ for $k_{m\rightarrow w}{=}10$ presented in figure \ref{figure5} (a). In the case of $k_{m\rightarrow w}{=}10$, the MMT exhibits a minimum at roughly $\delta_\text{opt}{=}0{.}1$ (figure \ref{figure5} (b)), while a minimum at $\delta_\text{opt}{=}0{.}2$ emerges for the number of waiting periods and thus also for the MWT (figure \ref{figure5} (c) and inset). Hence, the optimal cortical width $\delta_\text{opt}$ of the MFPT, which is the sum of MMT and MWT, undergoes a crossover from $\delta_\text{opt}{=}0{.}1$ to $\delta_\text{opt}{=}0{.}2$ as the impact of waiting times increases with decreasing rate $k_{w\rightarrow m}$.\\

\noindent
In conclusion, the MWT dominates the behavior of the first passage properties in the limit of $k_{w\rightarrow m}\rightarrow 0$, which is biologically relevant as the mean waiting time per arrest state is typically of the order of seconds \cite{Zaliapin2005,Slepchenko2007,Balint2013}. This results in a shift of the optimal width $\delta_\text{opt}$ in dependence of the mesh size of the underlying network defined by $k_{m\rightarrow w}$. For small rates $k_{m\rightarrow w}$ a raising impact of waiting times may lead to a shift of $\delta_\text{opt}$ from intermediate values to $\delta_\text{opt}{=}1$ (compare to the inset of figure \ref{figure5} (c)) and thus a favour of homogeneous cytoskeleton. Larger rates $k_{m\rightarrow w}$, and hence biologically relevant mesh sizes, conserve the gain of search efficiency by inhomogeneous cytoskeletal organizations, since the number of waiting periods exhibits a minimum for $\delta \in (0;1)$ (inset of figure \ref{figure5} (c)).

\begin{figure*}[t]
\centering
\includegraphics[width=0.97\linewidth]{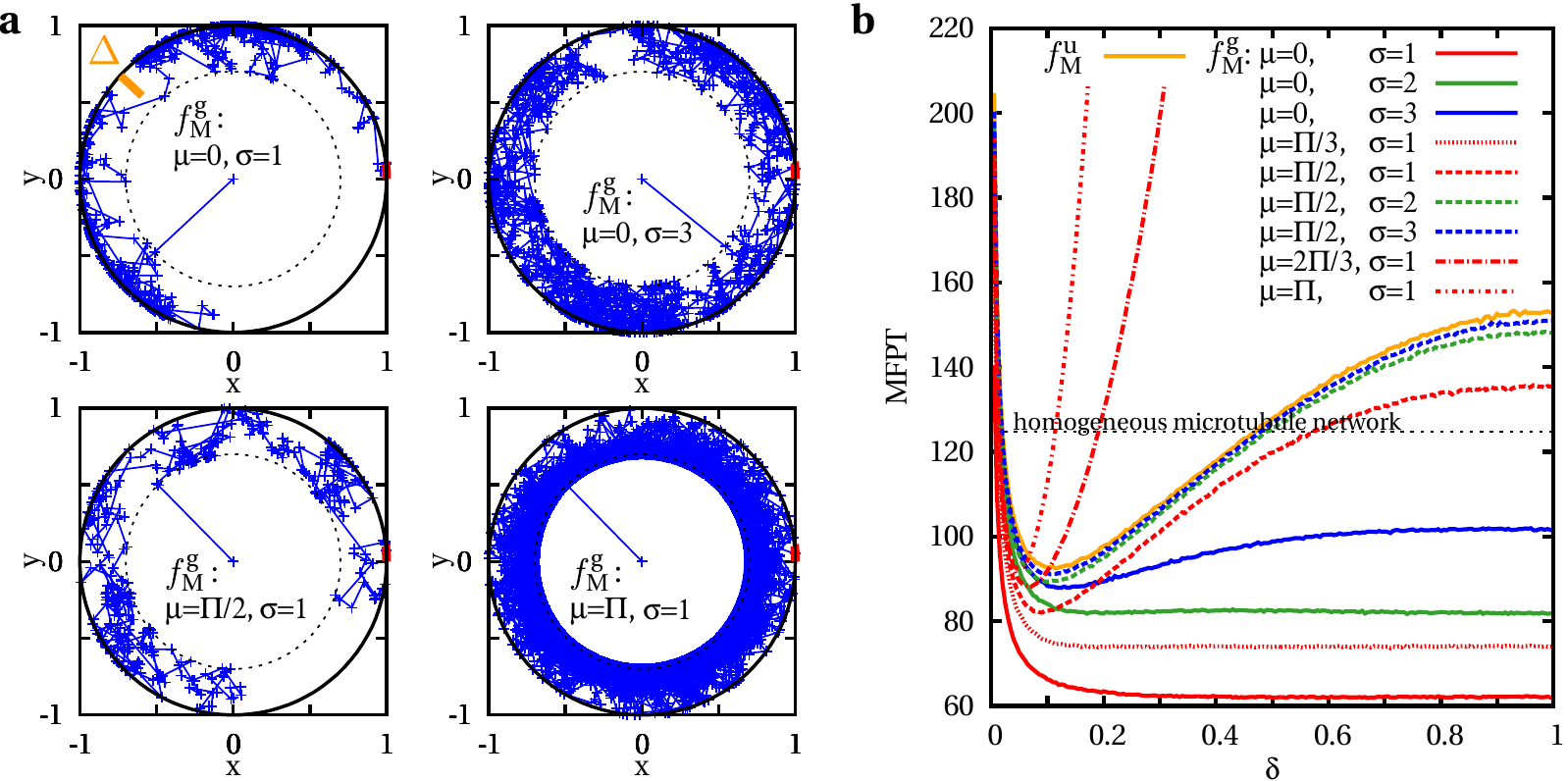}
\caption
{Influence of the Rotation Angle Distribution of the Actin Cortex on Inhomogeneous Search Strategies. 
Interior parameters: $k^i_{m\rightarrow w}{=}0$, $k^i_{w\rightarrow m}{=}\infty$; periphery parameters: $k^p_{m\rightarrow w}{=}10$, $k^p_{w\rightarrow m}{=}\infty$, $p{=}q_\text{K}{=}q_\text{D}{=}0$, $q_\text{M}{=}1$; target size: $\alpha_\text{target}{=}0{.}1$. \textbf{a} Sample trajectories for $\delta{=}0{.}3$ and various values of the rotation angle distribution of actin filaments. \textbf{b} MFPT in dependence of the actin cortical width $\delta$ for various values of the rotation angle distribution of actin filaments $f_\text{M}$.
}
\label{figure6}
\end{figure*}

\subsubsection*{Influence of the Rotation Angle Distribution of the Actin Cortex}

\noindent
The polarity of the actin filaments in the cortex is typically random. Hence we have previously investigated uniform rotation angle distributions $f_\text{M}^\text{u}$. \\
However, actin filaments may align to the radial microtubule network \cite{Lopez2014} and for instance the protein complex Arp2/3 induces a formation of actin branches at a distinct angle ($\approx 70^\circ$) compared to the parent filament \cite{Mullins1998,Risca2012}. In general, such mechanisms influence the orientational distribution of the actin filaments. \\

\noindent
Here, we study the impact of the expectation value and width of a cut-off-gaussian rotation angle distribution $f_\text{M}^\text{g}$ (as given in equation \ref{eq:rotanglemyosin}) on the search efficiency to narrow membranous targets. A mean value $\mu\!\in\![0;\pi/2)$ ($\mu\!\in\!(\pi/2;\pi]$) leads to outwardly (inwardly) peaked actin filaments, whereas $\mu{=}\pi/2$ corresponds to a predominantly lateral orientation. In order to isolate the influence of the orientational distribution we assume $k^i_{m\rightarrow w}{=}0$, $k^i_{w\rightarrow m}{=}\infty$ and $k^p_{m\rightarrow w}{=}10$, $k^p_{w\rightarrow m}{=}\infty$. Furthermore, we fix $p{=}0$, $q_\text{M}{=}1$ and $\alpha_\text{target}{=}0{.}1$.\\

\noindent
Figure \ref{figure6} displays the MFPT in dependence of the cortex width $\delta$, which defines the inhomogeneity of the system, for various actin orientations defined by the uniform distribution $f^u_\text{M}$ or the gaussian distribution $f^g_\text{M}(\mu,\sigma)$. Please first focus on $\sigma{=}1$, which corresponds to strongly peaked gaussian distributions $f^g_\text{M}(\mu,\sigma)$ of the actin filaments into directions defined by $\mu$.\\

\noindent
Inward-directed actin polarities ($\mu\!>\!\pi/2$) drive the cargo towards the cell center, which lowers the probability to reach the membrane and detect the target. As evident from the sample trajectory in figure \ref{figure6}, decreasing the actin cortical width $\delta$ draws the particle nearer to the cell's boundary. Consequently, a thin actin cortex is essential in order to improve the search efficiency to membranous targets as manifested by the prominent minimum of the MFPT for $\mu\!>\!\pi/2$ in figure \ref{figure6}. Lateral actin orientations ($\mu{=}\pi/2$) result in an equal chance for outward- and inward-directed motion of the cargo. Hence the behavior of the MFPT is similar to the one of uniformly random actin networks defined by $f^u_\text{M}$ and an inhomogeneous cytoskeletal structure with a thin actin cortex $\delta$ generally advances the search efficiency.
To the contrary, for outward-pointing actin filaments ($\mu<\pi/2$) the particle is predominantly moving in close proximity $\Delta$ of the cell's boundary, as sketched for the sample trajectory in figure \ref{figure6} (a). For that reason the cortex width $\delta$ does not significantly influence the MFPT as long as it is out of range of $\Delta$ and the homogeneous limit $\delta{=}1$ is most efficient, as evident from figure \ref{figure6}. This limit is even more efficient than a homogeneous microtubule network due to the outward-directed network structure, as found in figure \ref{figure2}. In general, a larger standard deviation $\sigma$ broadens the distribution of actin filaments and randomizes the search. Consequently, the behavior of the MFPT converges to the uniform case $f^u_\text{M}$ for increasing $\sigma$ and an inhomogeneous cytoskeleton generally improves the search of small membranous targets again.

\begin{figure*}[t]
\centering
\includegraphics[width=\linewidth]{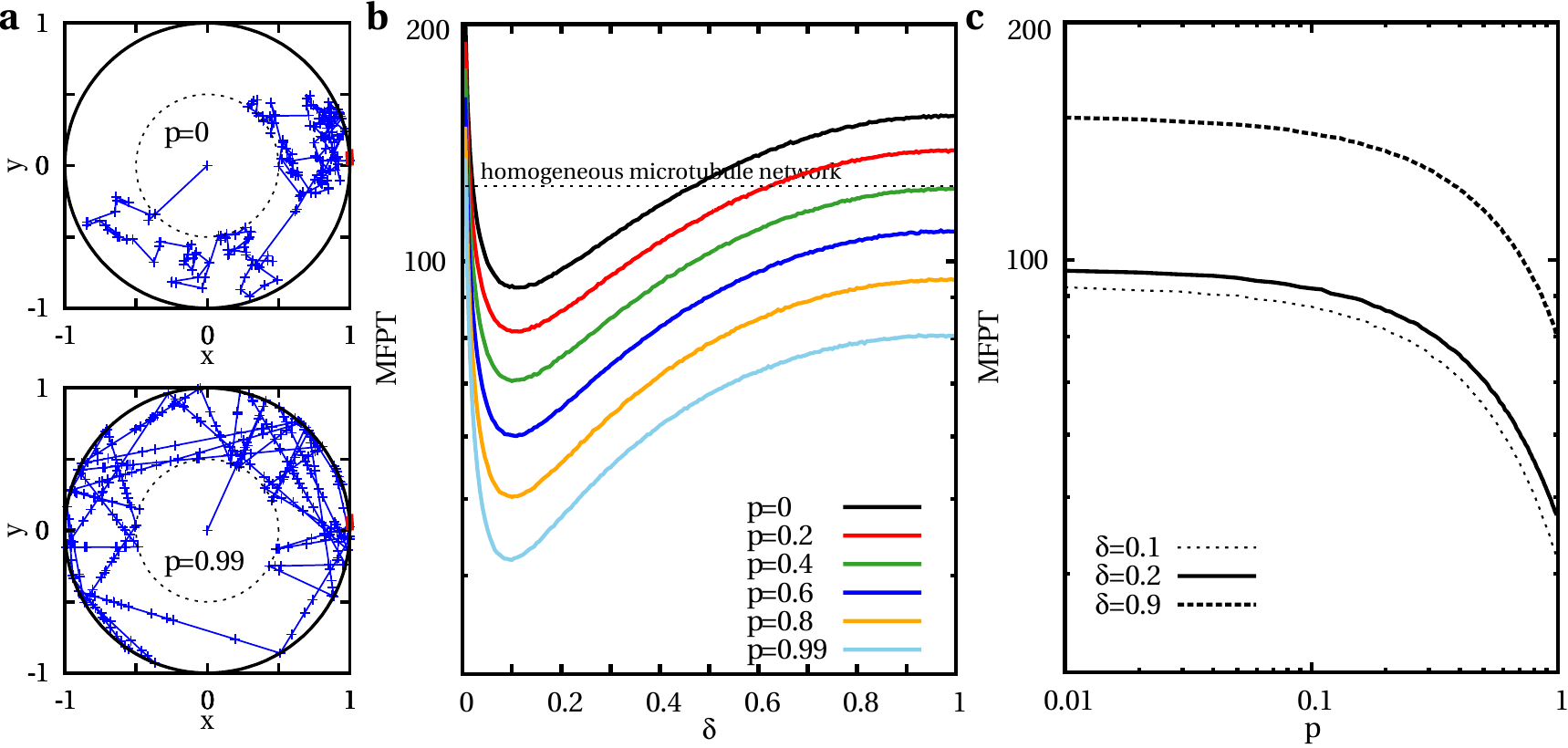}
\caption
{Influence of the Processivity on Inhomogeneous Search Strategies without Waiting Times. 
Interior parameters: $k^i_{m\rightarrow w}{=}0$, $k^i_{w\rightarrow m}{=}\infty$; periphery parameters: $k^p_{m\rightarrow w}{=}10$, $k^p_{w\rightarrow m}{=}\infty$, $q_\text{M}{=}1$; target size: $\alpha_\text{target}{=}0{.}1$. The crosses refer to intersection nodes. \textbf{a} Sample trajectories for $\delta{=}0{.}5$ and different values of the processivity $p$. \textbf{b} MFPT in dependence of the actin cortical width $\delta$ for diverse motor processivities $p$. An increase of the processive behavior at intersection nodes leads to a systematic decrease of the MFPT. Hence, purely processive motors are most efficient in the case of instantaneous directional changes. \textbf{c} The MFPT in dependence of the processivity $p$ exhibits a monotonic decrease for different values of the actin cortical width $\delta$. This emphasizes the benefit of highly processive motors.
}
\label{figure7}
\end{figure*}

\begin{figure*}[!htb]
\centering
\includegraphics[width=\linewidth]{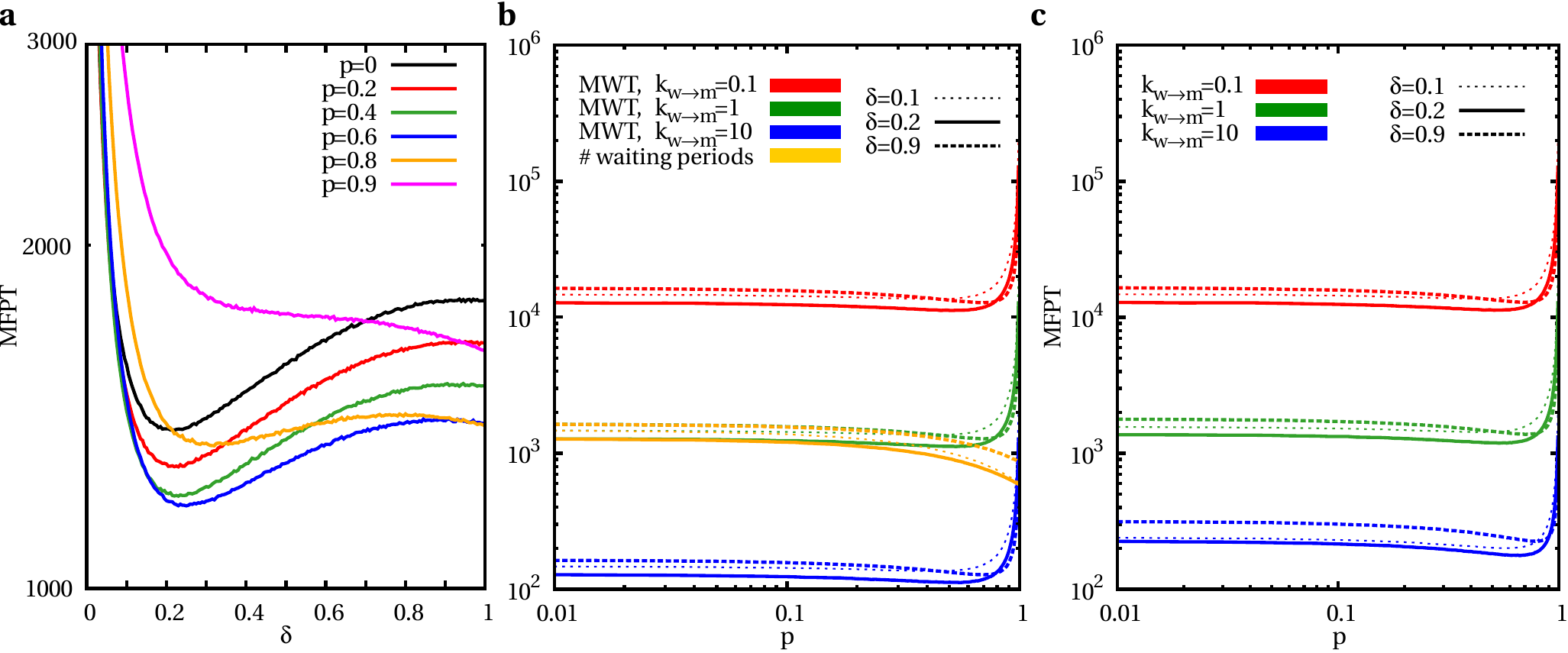}
\caption
{Influence of the Processivity on Inhomogeneous Search Strategies with Waiting Times. 
Interior parameters: $k^i_{m\rightarrow w}{=}0$, $k^i_{w\rightarrow m}{=}k_{w\rightarrow m}$; periphery parameters: $k^p_{m\rightarrow w}{=}10$, $k^p_{w\rightarrow m}{=}k_{w\rightarrow m}$, $q_\text{M}{=}1$; target size: $\alpha_\text{target}{=}0{.}1$. \textbf{a} MFPT in dependence of the actin cortical width $\delta$ for different motor processivities $p$ and $k_{w\rightarrow m}{=}1$. There arises a threshold at which a further increase of the processivity is disadvantageous. \textbf{b} The number of waiting periods decreases monotonically with the processivity $p$ for diverse fixed cortical widths $\delta$. Contrarily, the MWT diverges in the limit of $p\rightarrow 1$. \textbf{c} The MFPT in dependence of the processivity $p$ displays a minimum at $p\neq 1$ for various event rates $k_{w\rightarrow m}$ and different values of the actin cortical width $\delta$.
}
\label{figure8}
\end{figure*}

\subsubsection*{Influence of the Motor Processivity} 

\noindent
Molecular motors do not necessarily change their direction of motion at each intersection node they encounter. A main feature of molecular motors is their processivity. Motor driven cargoes may also overcome the barrier opposed by network intersections and keep moving ballistically along the same track, which has been neglected so far ($p{=}0$). Here, we would like to investigate the impact of the motor processivity $p$ on the search efficiency to narrow target zones. For that purpose, we assume $k^i_{m\rightarrow w}{=}0$, $k^i_{w\rightarrow m}{=}k_{w\rightarrow m}$ in the interior and a mesh size defined by $k^p_{m\rightarrow w} {=} 10$ in the periphery. After each waiting period at intersection nodes, the particle may either start moving processively at rate $p \, k_{w\rightarrow m}$ or it changes its direction at rate $(1-p) \, k_{w\rightarrow m}$. For simplicity, we fix $q_\text{M}{=}1$ and $\alpha_\text{target}{=}0{.}1$.

\noindent
The difference in the movement pattern by a change in processivity is visualized by sample trajectories presented in figure \ref{figure7} (a). figure \ref{figure7} (b) displays the MFPT in dependence of the actin cortical width $\delta$ for various values of the processivity $p$. By neglecting waiting processes at intersection nodes via $k_{w\rightarrow m}{=}\infty$, we find that a higher processivity systematically improves the search efficiency. This benefit is emphasized in figure \ref{figure7} (c). The MFPT for fixed actin cortical widths decreases monotonically in dependence on the processivity $p$ and is most efficient for $p{=}1$.\\

\noindent
Remarkably, the introduction of waiting processes results in the development of an optimal processivity $p\neq 1$, as presented in figure \ref{figure8} (a) for $k_{w\rightarrow m}{=}1$. For a fixed cortical width $\delta$, figure \ref{figure8} (b) shows that the mean number of waiting periods decreases with $p$ due to the overall gain in search efficiency by an enhanced processivity $p$. However the MWT drastically increases, since the mean waiting time at confinement events $[(1-p)\,k_{w\rightarrow m}]^{-1}$ diverges in the limit of $p\rightarrow 1$. figure \ref{figure8} (c) visualizes, that the MFPT is dominated by the MWT for small rates $k_{w\rightarrow m}$. Consequently, the MFPT exhibits a minimum for $p\neq 1$ in contrast to the MMT, which is minimal for $p{=}1$ (compare to figure \ref{figure7} (c)). Due to inevitable waiting processes, it may be more efficient to change directions with a specific probability $1{-}p$ rather than transport by completely processive motors ($p{=}1$). A specific processivity $p$ is also reported in biological systems \cite{Balint2013,Ali2007,Ross2008,Ross2008B}.

\FloatBarrier

\begin{figure*}[!htbp]
\includegraphics[width=0.88\linewidth]{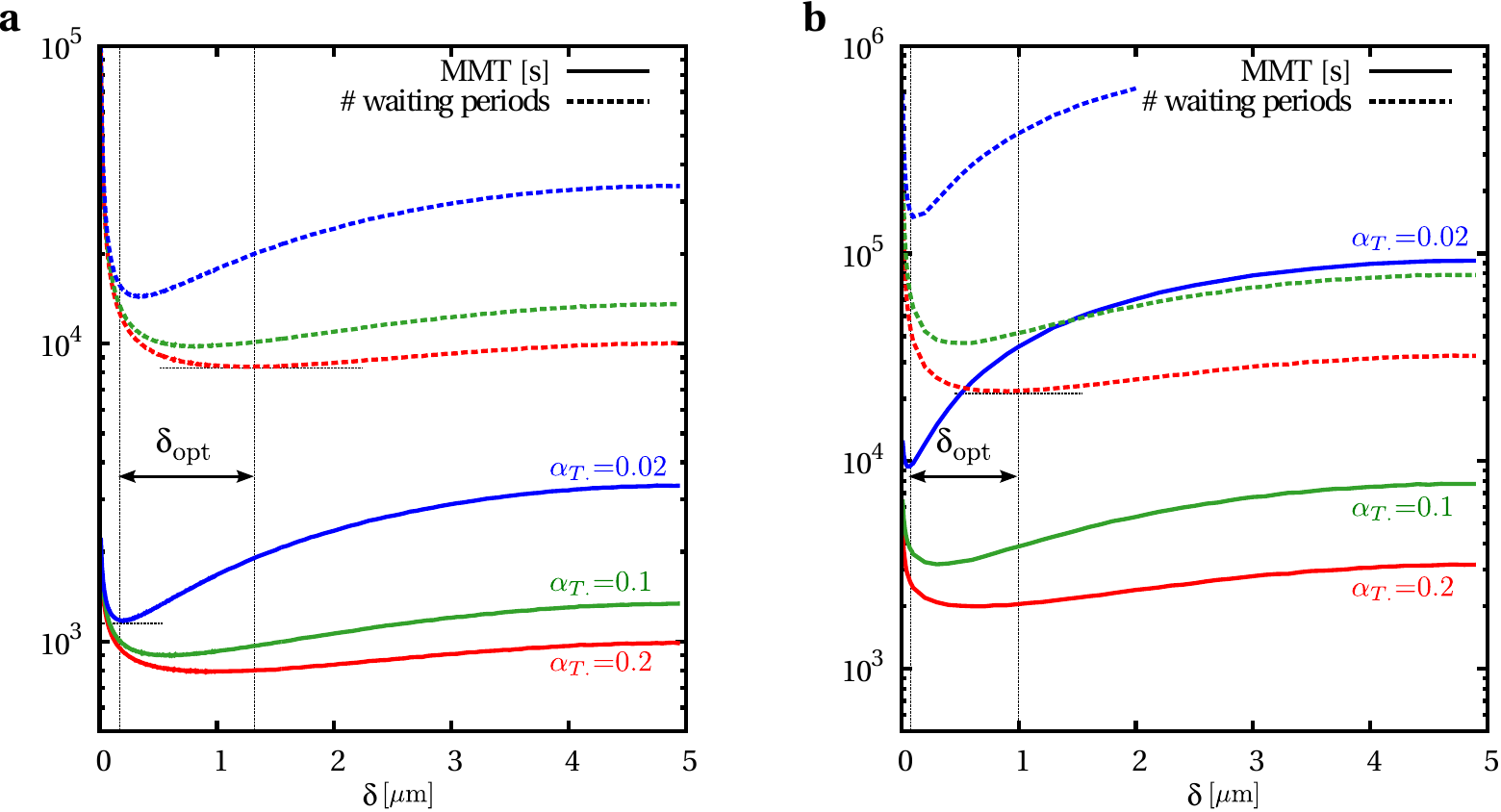}
\caption
{Comparison to Biological Dimensions. Interior Parameters: $k^i_{m\rightarrow w}{=}0/\text{s}$, $k^i_{w\rightarrow m}{=}k_{w\rightarrow m}$; periphery parameters: $k^p_{m\rightarrow w}{=}10/\text{s}$, $k^p_{w\rightarrow m}{=}k_{w\rightarrow m}$, $p{=}0{.}5$, $q_\text{M}{=}1$; velocity: $v{=}1\,\mu\text{m}/\text{s}$; search domain: $R_{m.}{=}5\,\mu\text{m}$; target size: $\alpha_\text{target}\in \{0{.}02,0{.}1,0{.}2\}$. The MMT (MFPT for $k_{w\rightarrow m}{=}\infty$) as well as the mean number of waiting periods display distinct minima at small cortical widths in dependence of the target size $\alpha_\text{target}$ for the 2D case (\textbf{a}) as well as the generalization to 3D (\textbf{b}). The optimal width of the actin cortex $\delta_\text{opt}$ varies according to $\alpha_\text{target}$ and the mean waiting time per arrest state $1/k_{w\rightarrow m}$ from approximately $\delta_\text{opt}{=}0{.}1$ $\mu$m to $\delta_\text{opt}{=}1{.}2$ $\mu$m in 2D and $\delta_\text{opt}{=}1$ $\mu$m in 3D, which is in qualitative agreement to biological measurements.
}
\label{figure9}
\end{figure*}

\vspace*{5cm}
\pagebreak
\vspace*{5cm}

\section*{Discussion}

\noindent
With the aid of a random velocity model with intermittent arrest states, we studied the first passage properties of intracellular narrow escape problems. Via extensive computer simulations we are able to systematically analyze the influence of the cytoskeletal structure as well as the motor performance on the search efficiency to small target zones on the plasma membrane of a cell.\\

\noindent
For a spatially homogeneous cytoskeleton, the MFPT diverges in the limit $\alpha_\text{target}\rightarrow 0$ and directional changes at intersection nodes ($k_{m\rightarrow w}\neq 0$, $p{=}0$) do not improve the search efficiency. Moreover, we find that a random actin mesh ($q_\text{M}{=}1$) is preferable to radial microtubule networks ($q_\text{M}{=}0$) as well as combinations of both ($q_\text{M} \in (0;1)$), which underlines the benefit of motor activity regulation by outer stimuli. Homogeneous motion pattern are sufficient for large target sizes, but actually fail in the biologically relevant case of narrow escape problems.\\

\noindent
By varying the width of the actin cortical layer, we elaborate the impact of the cytoskeletal inhomogeneity on the search efficiency to narrow escapes. Remarkably, we find that a cell can optimize the detection time by regulation of the motor performance and convenient alterations of the spatial organization of the cytoskeleton. An inhomogeneous architecture with a thin actin cortex constitutes an efficient intracellular search strategy for narrow targets and generally leads to a considerable gain of efficiency in comparison to the homogeneous pendant. A confinement of the search to a thin shell below the plasma membrane, where the target area is located, saves time as it prevents extensive excursions to the cell interior - but the shell must not be too thin because otherwise the motion of the searcher gets localized and a time loss due to many stops occurs. Consequently, the MFPT diverges in the limit of $\delta \rightarrow 0$, which outlines that the actin cortex can act as a transport barrier or functional gateway \cite{Papadopulos2013}.\\

\noindent
Molecular motors do not necessarily change their direction of motion at each filamentous intersection, they may also overcome the constriction and remain on the same track, a property referred to as processivity. We find that an increased motor processivity systematically improves the search efficiency in the case of instantaneous directional changes on networks of a mesh size defined by $k^p_{m\rightarrow w}$. Due to waiting processes an optimal value of the motor processivity $p\neq 1$ emerges, which minimizes the detection of membranous targets. Specific probabilities to overcome constricting filament crossings are also reported during intracellular transport \cite{Balint2013,Ali2007,Ross2008,Ross2008B}.\\

\noindent
So far we have investigated a model for intracellular search strategies in two-dimensional cells. However, a generalization of our approach to three-dimensional spherical cells is straightforward and the former principles stay valid. If we assume a cell radius of $5$ $\mu$m, which is consistent with the typical size of a CTL, the opening angle of $\alpha_\text{target}{=}0{.}02$ ($\alpha_\text{target}{=}0{.}2$) leads to an arc length of $0{.}02\times 5$ $\mu$m ${=}100$ nm ($1$ $\mu$m). For instance the diameter of an immunological synapse is of the order of microns \cite{Grakoui1999,Qu2011,Angus2013,Ritter2013}. We further assume that motors move processively at intersections with probability $p{=}0{.}5$ and their velocity typically is about $1$ $\mu$m$/$s. A transition rate to the waiting state of $10/$s thus leads to a mesh size of $100$ nm, which is biologically reasonable \cite{Eghiaian2015,Salbreux2012}. Under these conditions, figure \ref{figure9} reveals an optimal width of the actin cortex which varies from approximately $\delta_\text{opt}=0{.}1$ $\mu$m to $\delta_\text{opt}=1{.}2$ $\mu$m in 2D and $\delta_\text{opt}=1$ $\mu$m in 3D. This is in good qualitative agreement to biological data \cite{Eghiaian2015,Salbreux2012,Clark2013}.\\

\noindent
In summary, our model indicates that the spatial organization of the cytoskeleton of spherical cells with a centrosome minimizes the characteristic time necessary to detect small targets on the cell membrane by random intermittent search processes along the cytoskeletal filaments (see also \cite{Schwarz2016,Schwarz2016b} for similar findings in a model with intermittent diffusive search). The minimization is achieved  by a small width of the actin cortical layer and by regulation of the motor activity and behavior at network intersections. Remarkably a thin actin cortex is also more economic than distributing cytoskeletal filaments in all directions over the cell body. Thus, our work outlines that a thin confinement of the actin cortex, besides its advantages concerning cell stability or motility, also serves as an efficient key to intracellular narrow escape problems.

\section*{Acknowledgements}

\noindent
This work was funded by the German Research Foundation (DFG) within the Collaborative Research Center SFB 1027.

\end{document}